\documentclass[conference]{IEEEtran}
\IEEEoverridecommandlockouts
\usepackage{cite}
\usepackage{amsmath,amssymb,amsfonts}
\usepackage{algorithmic}
\usepackage{graphicx}
\usepackage{hyperref}
\usepackage{multirow}
\usepackage{textcomp}
\usepackage{bm}
\usepackage{xcolor}
\def\BibTeX{{\rm B\kern-.05em{\sc i\kern-.025em b}\kern-.08em
    T\kern-.1667em\lower.7ex\hbox{E}\kern-.125emX}}
\begin{document}

\title{Velocity Prediction in Automatic Guitar Transcription\\
\thanks{This work is supported by Innovate UK (Project no. 10105279) and the UKRI Centre for Doctoral Training in Artificial Intelligence and Music (Grant no. EP/S022694/1). This research utilised Queen Mary's Apocrita HPC facility, supported by QMUL Research-IT. \href{http://doi.org/10.5281/zenodo.438045}{http://doi.org/10.5281/zenodo.438045}}
}


\author{\IEEEauthorblockN{Jackson Loth, Xavier Riley, Simon Dixon and Emmanouil Benetos}\IEEEauthorblockA{Centre for Digital Music\\
Queen Mary University of London, United Kingdom}}

\maketitle

\begin{abstract}
Automatic Music Transcription (AMT) models have achieved a high level of success in polyphonic transcription of various instruments. Velocity, typically a measure of note intensity, is less commonly predicted in these models due to the absence of velocity labels in available datasets and lack of a proper definition for instruments other than piano. We present a methodology and model for velocity prediction in Automatic Guitar Transcription (AGT) which uses virtual instruments to generate synthetic training data with velocity labels. We first pretrain a model on this synthetic data. These weights are then transferred to a different model and trained on real guitar audio, allowing the model to retain the working velocity prediction while also achieving high performance and generalisability from the real training data. The velocity prediction is shown to outperform a baseline model which does not use the pretrained velocity weights, when evaluated on synthetic data. In addition, using the pretrained velocity weights offers a small improvement in note transcription, though the magnitude of this improvement is limited and not always significant depending on the testing data. Overall the model achieves results comparable to the state of the art in guitar transcription, while also successfully predicting velocity.
\end{abstract}


\section{Introduction}

Automatic Music Transcription (AMT) involves transcribing musical audio to some symbolic format such as MIDI \cite{benetos2018automatic}. Like many music information retrieval (MIR) tasks, it has found great success in recent years through the use of deep learning. While research has explored a variety of different instruments, a small dedicated subfield has grown around the guitar. Commonly referred to as Automatic Guitar Transcription (AGT), progress in this task has been steadily increasing in recent years. However, focusing on guitar can be challenging due to the relative difficulty of obtaining high-quality data compared to other instruments such as piano. To address this, several datasets of guitar audio have been created and made available by researchers, such as GuitarSet\cite{xi_guitarset_2018}, IDMT-SMT-GUITAR\cite{kehling_2014_automatic}, EGDB \cite{chen2022towards}, GAPS \cite{riley2024gaps}, Guitar-TECHS \cite{pedroza2025guitar} and GOAT \cite{loth2025goat}. These datasets all provide note-level and sometimes string-level annotations with real guitar audio, but are also unfortunately difficult and time-consuming to create. This difficulty can however be offset somewhat through score alignment techniques \cite{maman2022unaligned,riley2024high,yaffe2025count} by automatically aligning scores with audio, allowing annotations to be more easily created without specialised hardware such as hexaphonic pickups.

While these works achieve impressive results in pitch transcription, velocity tends to be ignored by AGT papers. Velocity is a somewhat ambiguous and abstract value in the context of guitar, despite being an important component of expressiveness in a musical performance. While it is generally thought of as an indication of dynamic level or loudness, it tends to have a nonlinear relationship with loudness \cite{dannenberg2006interpretation} and is perhaps better thought of as a general measure of ``intensity", related to both loudness and timbre. The values that are controlled by velocity in virtual instruments are left more to the developer of the instrument rather than following any kind of standard. Given this vague definition of velocity, it is unsurprising that, outside of piano transcription where velocity is actually measured with a Disklavier \cite{hawthorne2017onsets,kong}, velocity is often ignored in favor of focusing on transcribing each note's pitch, onset and offset. Datasets such as MAESTRO \cite{hawthorne2018enabling}, Studio MAESTRO \cite{edwards2024data} and Saarland \cite{muller2011saarland}, which were created using a Disklavier, enable easier training and evaluation of piano velocity transcription. While DSP-based approaches to velocity prediction exist \cite{ewert2011estimating}, most recent papers focus on deep learning-based approaches  \cite{kim2023score,kim2023diffvel,kim2024method,jeong2018timbre,jeong2020note}.

Given these difficulties with velocity and the absence of any datasets which contain proper velocity annotations for guitar, we instead use the velocities defined by guitar virtual instruments to create a synthetic dataset which we use to train our model. These velocity labels, which broadly map to volume and timbre in the virtual instrument, are used to define velocity in the context of this model. Once we train this initial velocity model, we can integrate the trained velocity prediction into a separate model trained on real-world audio, giving us the velocity prediction along with the accuracy of a model trained on high quality real data. The approach of using synthetic data to train AGT models has been shown to be effective in note onset and offset prediction \cite{kusaka2025learn}, though this approach has not yet been applied to guitar velocity prediction. Prior work has however shown promise in a similar approach for predicting MIDI expression values in wind and string instruments \cite{xie2025finegrained}.

The contents of this paper are as follows: (1) an overview of the data collection, training and evaluation methodology; (2) an evaluation of the velocity prediction; and (3) an evaluation of the effects of velocity prediction on the note transcription. To our knowledge this is the first work to extend guitar transcription to also include velocity prediction.
\begin{figure*}
    \centering
    \includegraphics[width=\textwidth]{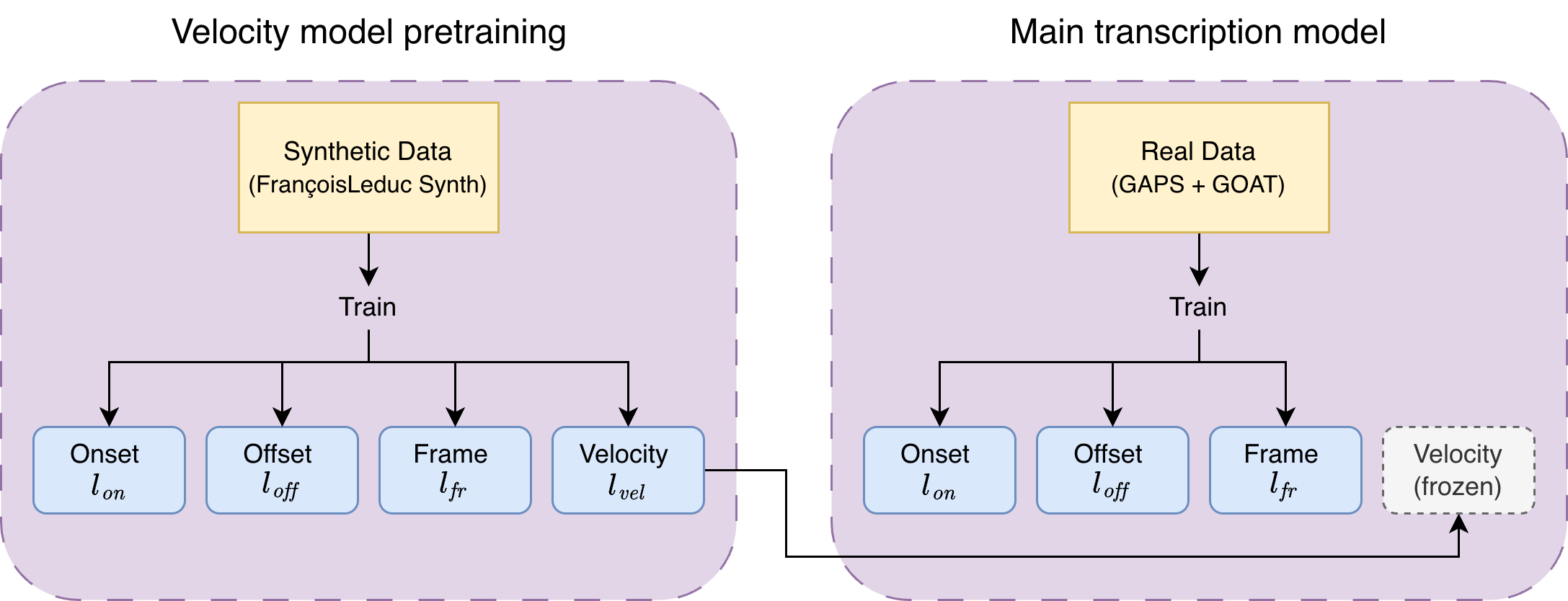}
    \caption{Visualization of the training methodology.}
    \label{fig:vel_diagram}
\end{figure*}

\section{Methodology}

\subsection{Data}

To train the velocity prediction part of our model, we use a synthetic dataset. While publicly available synthetic datasets of guitar audio such as SynthTab \cite{zang2024synthtab} exist, they do not contain suitable velocity annotations. Our data was instead created using the FrançoisLeduc dataset \cite{riley2024high}, a collection of 79 aligned real guitar audio and MIDI annotations spanning four hours.

For each note in the MIDI annotations, we calculated a rough estimate of the velocity.
The estimated velocities were obtained by first isolating the audio corresponding to the onset and offset of each note. For each note segment, the audio was transformed into a frequency representation and filtered using bandpass filters centered at the note's fundamental frequency and harmonics. We then computed the total energy of each filtered signal and applied a weighted sum with exponentially decaying weights for each harmonic. This was normalised to a range of 0 to 127 to convert to velocity values.

This estimated velocity information was applied to the corresponding MIDI annotations and rendered using five different timbres from three guitar virtual instruments \footnote{\href{https://www.native-instruments.com/en/products/komplete/guitar/session-guitarist-picked-nylon/}{Native Instruments Session Guitarist: Picked Nylon};\\\href{https://www.native-instruments.com/en/products/komplete/guitar/session-guitarist-picked-acoustic/}{Native Instruments Session Guitarist: Picked Acoustic};\\\href{https://www.straightaheadsamples.com/smart-delay-2}{Straight Ahead Samples JB-145: Archtop Guitar}}, creating roughly 20 hours of synthetic training data with velocity.
Through this process, we were able to obtain synthetic guitar audio in which the performances correspond exactly to the velocity values in the performance MIDI. The velocity curve used by these virtual instruments (the level/timbre/intensity which corresponds to each velocity value) defines the ``true" velocity curve which our models learn to predict.

In addition, we use a variety of real audio datasets in training and evaluation. The GAPS \cite{riley2024gaps} and GOAT \cite{loth2025goat} datasets contain large amounts (14 hours and 6 hours respectively) of high quality real guitar recordings from two distinct guitar timbres (nylon string and electric guitars respectively). These two datasets are used to fine-tune our model on real guitar data. We also utilize GuitarSet \cite{xi_guitarset_2018} and EGDB \cite{chen2022towards} for testing. These datasets similarly provide two distinct types of guitar timbres (3 hours of acoustic and 2 hours of electric guitar respectively). The velocity parameter in these datasets is typically set to some constant value. In the case of both GAPS and GOAT, this value is 100.

\subsection{Model and Training}

We adopt the High Resolution Piano Transcription model proposed by Kong et al. \cite{kong}, itself having also been used in several AGT papers \cite{riley2024high,riley2024gaps,loth2025goat} to achieve state-of-the-art results. This model is a convolutional recurrent neural network (CRNN) which is trained to predict onset, offset, velocity and frame activation for each pitch across 10-second log mel-spectrogram inputs. The predicted velocity values range from 0 to 127, as in the MIDI standard. The model architecture is composed of several submodules (onset, offset, frame and velocity), with each comprising of several convolutional layers, bidirectional gated recurrent units and a fully connected layer. During training, each submodule calculates a binary cross entropy loss (onset $l_\mathit{on}$, offset $l_{\mathit{off}}$, frame $l_\mathit{fr}$ and velocity $l_\mathit{vel}$), of which the summation of each is used as the main loss function (see Equation \ref{eq:loss_A}). We scale the offset loss by a factor of 0.1 in order to reduce the effect of imprecise offset labels in the training data, as done in \cite{riley2024high}. 


\begin{equation}
\label{eq:loss_A}
    \mathcal{L_A} = l_\mathit{on} + (0.1)l_{\mathit{off}} + l_\mathit{fr} + l_\mathit{vel}
\end{equation}

We use two training stages to train our model. The goal of the first stage is just to learn the velocity curve of the synthetic data. In the second stage, our goal is to train the rest of the submodules on real data so that the note onset/offset transcription system works well on realistic data. We then combine the trained velocity submodule from the first stage with the rest of the trained submodules from the second stage. See Figure \ref{fig:vel_diagram} for a visual representation of this training methodology.

For the first stage of training, we first trained a model on the synthetic FrançoisLeduc dataset. We use loss $\mathcal{L_A}$ to train this model. Once this has been trained and the velocity curve of the synthetic data has been learned, we move to the second stage and train on our real datasets. The weights from the pretrained velocity model component are loaded and frozen in order to prevent the velocity prediction model from changing during training, as the training data will no longer have correct velocity labels. The model is then trained using the GAPS and GOAT datasets using a modified loss $\mathcal{L}_B$. This simply removes the velocity term entirely from the loss. As our datasets of real guitar audio do not have correct velocity labels, this change to the loss function prevents these labels from having any effect on the overall calculated loss. This allows us to have both a high quality and generalizable note prediction model while still having a working velocity prediction model.

\begin{equation}
\label{eq:loss_B}
    \mathcal{L}_B = l_\mathit{on} + (0.1)l_{\mathit{off}} + l_\mathit{fr}
\end{equation}

A few different types of data augmentation are applied to the data during training. First, two peak filters are applied to a random frequency between 32 and 4096 Hz, with Q randomly set between 1 and 2 and gain randomly set between -30 dB and 10 dB. Next, a reverb is applied with the mix randomly set between 0\%  and 70\% wet. Finally, a random impulse response, recorded from several different microphone responses, is applied 50\% of the time. These augmentations are applied to the signal using the Pedalboard package \cite{sobot_peter_2023_7817838}.

All models were trained from a pretrained piano transcription checkpoint, as done in \cite{riley2024high}. During the second stage of training, we initially tried fine-tuning from the checkpoint of the model trained on synthetic data in the initial stage of training. However, we found this to cause the model to overfit extremely quickly before seeing all of the training data. The models were trained using a single NVIDIA A100 GPU. When pretraining on the synthetic data to learn the velocity prediction, the model was trained for $150000$ iterations using a learning rate of $0.0001$ and a batch size of six. When training on real data, the learning rate was changed to $0.00001$. The performance generally plateaued after around 20k iterations in all models.
\section{Results}

\subsection{Velocity Prediction}
\label{sec:velocity_results}

Unlike piano, there are no existing guitar datasets with both ground truth velocity annotations and real (non-synthetic) audio known to the authors. As previously discussed, the concept of velocity is also difficult to apply to guitar as it is not an actual measured value in instruments other than piano, but a more abstract measure of intensity related to both note volume and timbre. In this work we relied on the velocity curves defined by the virtual instruments used to render the synthetic guitar training dataset. Because of this, any evaluation would depend on how closely the velocity annotations in the test data matched the implementation of velocity in the virtual instruments used for training data, making it possible for our model to not generalise to other implementations even if it has properly learned the velocity curve of its training data. To handle this, we train two separate models using our velocity methodology, which each use a different split of the data (which we refer to as the \emph{song} and \emph{timbre} data splits), along with two baselines in each data split which do not incorporate our velocity prediction method. The first model, using the \emph{song} data split, is trained on all five timbres. The test data is only taken from songs which do not appear in the training data. The second model, using the \emph{timbre} data split, is only trained on the JB-145 timbres, while the Native Instrument timbres are used as the test data. This allows us to also observe how our velocity prediction generalizes to an unknown timbre.

All of our models are fine-tuned using the GAPS and GOAT datasets. To evaluate the velocity prediction itself we use the mean absolute error, as proposed by Kim and Serra \cite{kim2024method} (Equation \ref{eq:vel_error}), where $V(i)_\mathit{gt}$ and $V(i)_\mathit{est}$ represent the velocity of note $i$ of the ground truth and estimated notes respectively and $N$ represents the total number of correctly matched notes in the score. 

\begin{equation}
\label{eq:vel_error}
    \mathit{Error} = \frac{\sum_i|V(i)_\mathit{gt}-V(i)_\mathit{est}|}{N}
\end{equation}

This is calculated over all matched pairs of MIDI notes. For evaluating automatic transcription performance, we also calculate F1 scores, using the matched MIDI notes while also only considering two notes matched if the velocities are within a threshold of $0.1$ of each other after scaling to $[0,1]$. The F1 score are also calculated without velocity as a comparison. These calculations were performed using the \verb|mir_eval|\footnote{\href{https://mir-eval.readthedocs.io/latest/api/transcription_velocity.html}{https://mir-eval.readthedocs.io/latest/api/transcription\_velocity.html}} package \cite{raffel2014mireval} and are presented in Table \ref{tab:vel_results}. We include both the mean and standard deviation of the error scores (Equation \ref{eq:vel_error}).

\begin{table}
    \centering
    \caption{Velocity transcription results on the synthetic FrançoisLeduc dataset. Models are trained using two different train/test splits of the dataset. Error is calculated using Equation~\ref{eq:vel_error}. SD represents standard deviation. $F_{50}$ is F1-measure at 50\,ms resolution, while $F_{50}$ (vel) additionally accounts for velocity prediction accuracy. Best metrics are shown in bold.}
    \resizebox{\columnwidth}{!}{%
    \begin{tabular}{|c|c|cccc|}
        \hline
        \multicolumn{2}{|c|}{Model} & \multicolumn{4}{c|}{} \\
        \hline
        Data split & Method
        & $\mathit{Error}$ (mean) $\downarrow$
        & $\mathit{Error}$ (SD) $\downarrow$
        & $F_{50}$ (vel) $\uparrow$
        & $F_{50}\uparrow$ \\
        \hline

        \multirow{2}{*}{\emph{song}} 
            & Baseline & $32.39$ & $19.78$ & $35.6$  & $91.19$ \\
            & Ours & $\bm{7.04}$ & $\bm{5.52}$ & $\bm{69.22}$ & $\bm{91.22}$ \\
        \hline

        \multirow{2}{*}{\emph{timbre}} 
            & Baseline & $33.59$ & $20.54$ & $34.7$  & $93.74$ \\
            & Ours & $\bm{11.53}$ & $\bm{9.49}$ & $\bm{51.99}$ & $\bm{94.14}$ \\
        \hline
    \end{tabular}
    }
    \label{tab:vel_results}
\end{table}

The model trained for guitar velocity prediction unsurprisingly obtained significantly better scores than the one without such training in both test splits. Anecdotally, transcribing unseen guitar recordings seems to indicate that the model generalizes well to real data, indicating the intensity of a played note in a useful way. Unfortunately this is difficult to test definitively, due to the lack of ground truth velocity annotations.

\subsection{Note Onset and Offset Prediction}
\label{sec:note_results}
Velocity prediction has been shown to help improve note prediction scores \cite{kong}, as the note prediction is conditioned on the velocity through concatenation with an intermediate output of the onset prediction module. This allows the model to attend more to the frame when the velocity is low. In previous papers using the same transcription architecture and pretraining \cite{riley2024high,riley2024gaps,loth2025goat}, velocity prediction was still used in the model, despite the fact that the fine-tuning data's (GAPS, GOAT) velocity labels were constant values. Having demonstrated the velocity prediction in Section \ref{sec:velocity_results}, we now use our model to test whether trained velocity prediction offers any benefit over the previous approach. We trained models in the same way as for the experiment in Section \ref{sec:velocity_results} for both our proposed model and the baseline model. We use both GuitarSet and EGDB datasets in order to test results on both acoustic (GuitarSet) and electric (EGDB) guitar timbres, timbres which are present in both of the training datasets. While training the models for this experiment, we noticed that the differences between the evaluation metrics fluctuated more than expected between training runs. To control for this, we trained three models in each category and averaged each run together for the final evaluation metrics. We present precision, recall and F1 scores, though velocity prediction is excluded from this calculation.

\begin{table}[htbp]
    \centering
    \caption{Transcription results over all of GuitarSet. $P_{50}$, $R_{50}$ and $F_{50}$ are precision, recall and F1-measure at 50ms resolution. Best metrics marked in bold.}
    \begin{tabular}{|c|c|c|c|}
        \hline
          & $P_{50}\uparrow$ &  $R_{50}\uparrow$ & $F_{50}\uparrow$\\
         \hline
         Baseline & $90.36$  & $83.94$  & $86.77$  \\
         Proposed model & $\bm{90.45}$ & $\bm{84.04}$ & $\bm{86.87}$ \\
         \hline
    \end{tabular}
    \label{tab:note_gs_results}
\end{table}

\begin{table}[htbp]
    \centering
    \caption{Transcription results over all of EGDB. $P_{50}$, $R_{50}$ and $F_{50}$ are precision, recall and F1-measure at 50ms resolution. Best metrics marked in bold.}
    \begin{tabular}{|c|c|c|c|}
        \hline
          & $P_{50}\uparrow$ &  $R_{50}\uparrow$ & $F_{50}\uparrow$\\
         \hline
         Baseline & $83.92$ & $83.04$ & $82.86$\\
         Proposed model & $\bm{84.05}$  & $\bm{83.07}$ & $\bm{82.95}$\\
         \hline
    \end{tabular}
    \label{tab:note_egdb_results}
\end{table}

Tables \ref{tab:note_gs_results} and \ref{tab:note_egdb_results} present the results of the models when tested on all of GuitarSet and EGDB respectively. Using the pretrained velocity weights does give a slight boost in performance in all metrics when evaluated on both GuitarSet and EGDB. For GuitarSet these differences were found to be statistically significant using a paired $t$-test with $\alpha = 0.05$ (precision: $p=0.006$, recall: $p=0.016$, F1: $p<0.001$ for GuitarSet), with Shapiro-Wilk tests indicating normality in the distribution of differences across the test datasets. When tested on the EGDB dataset, the differences in precision scores were found to be statistically significant ($p=0.027$), again with normality indicated using Shapiro-Wilk tests. However, differences in recall and F1 scores were not found to be statistically significant for EGDB.

\section{Discussion}

Using our proposed method for velocity prediction is shown to be successful when tested on the test splits of the synthetic dataset. Given that velocity lacks a precise definition in the MIDI protocol, our goal with velocity prediction is more about getting a good relative measure of the intensity of a note rather than a precise value. 
Regarding the song data split, a mean error of $7.04$, along with a standard deviation of $5.52$, seems to indicate that the model is predicting along the velocity curve generally as intended. An error of 7 is less than 6\% of the total range of 128 values, which may be acceptable for many use cases. For comparison, it is worth considering that Western music notation rarely uses more than a handful of dynamics markings (ppp, pp, p, mp, mf, f, ff, fff), much less than the number of levels that our approach can reliably distinguish. 

Comparing the results of the different data splits, we observe that most of the metrics became worse in the timbre split model than the song split model. However, the difference between our model and the baseline model in the timbre data split remains large. Despite having a higher mean and SD error compared to song split, the timbre split results still indicate that the model is adequately predicting the velocity of the notes, with the mean error of 11.53 still being only roughly 9\% of the total velocity range of 128. This is a good indication that the prediction generalises to different timbres. The worsening metrics are likely due to the differences in how the velocities are defined in the Native Instruments virtual instruments.

The $F_{50}$ (vel) results are unsurprisingly lower than the regular $F_{50}$ scores. This is in line with what we would expect given the way the algorithm uses the 10\% threshold for velocity (corresponding to 12.8 in the 0-127 velocity scale). The decrease is nearly double in the case of the timbre data split (roughly 42\% vs 23\% decrease in the song data split). This however is somewhat expected due to the larger error results from these models. Despite this decrease, these results are still promising, showing a clear increase compared to the baseline in both data splits.


The evaluation metrics in Tables \ref{tab:note_gs_results} and \ref{tab:note_egdb_results} show transcription results which are comparable to, though not surpassing, state-of-the-art performance in zero-shot evaluation on GuitarSet (see \cite{riley2024gaps}). Our proposed model improves on the baseline performance on the test datasets, although this improvement is only statistically significant in the case of GuitarSet. Regardless of statistical significance however, these boosts in performance are extremely small, around $0.1\%$. Due to the small magnitude of performance improvement, we suspect that other areas such as architectural changes or improvement or addition of data could have more potential for improving model performance.

An AGT model with working velocity prediction could have many uses to researchers. The predicted velocities could lead to more fine-grained automatic analysis of expressive musical performance, as well as help guitar modeling tasks due to the extra expressive information. The model also has potential in educational contexts, allowing guitarists to learn not only what notes are played in a given performance, but also more information about how the notes are played.

The biggest limitation of our methodology is the reliance on synthetic data and the inability to evaluate the velocity prediction on real guitar audio. Unfortunately this limitation stems from the fact that velocity is not a real measurable value in the context of guitar. It would be very beneficial for future work to address this issue, either by defining velocity in the context of guitar in terms of parameters of the excitation (how the string is plucked) or parameters of the resulting sound (energy or loudness), or through self-supervision, as suggested by Maman and Bermano \cite{maman2022unaligned}.
\section{Conclusion}

In this paper, we presented a training methodology for an AGT model which also transcribes velocity. Velocity is difficult to work with in guitar tasks due to the absence of available data and the inability to concretely measure velocity outside of instruments such as piano. To address this issue, we generated synthetic data using virtual instruments which define the velocity curve predicted by our model. We first trained a model on synthetic data, then transferred the weights from the velocity component over to a model which is trained on real guitar audio data. This velocity transcription method was found to be effective and generalise well when tested on the synthetic data. It was also found that using the pretrained velocity weights gave a slight improvement to onset/offset transcription metrics. However this improvement was not always significant depending on the data used to test the model, and the improvement itself was quite limited. 

Velocity transcription for guitar opens up many possibilities in MIR tasks related to guitar. We hope that this work is useful to other researchers in the field and that it inspires more work focusing on velocity in guitar transcription.




\bibliographystyle{IEEEtran}
\bibliography{main}

\end{document}